# IMPACT OF BUBBLE NUCLEATION ON THE FUNCTIONING OF THE PULSATING HEAT PIPE: NUMERICAL SIMULATION STUDY

**Vadim Nikolayev†,* and Iaroslav Nekrashevych**

*Service de Physique de l'Etat Condensé, CEA, CNRS, Université Paris-Saclay, CEA Saclay, 91191 Gif-sur-Yvette Cedex, France*

†Presenting Author: vadim.nikolayev@cea.fr
*Corresponding Author: vadim.nikolayev@cea.fr

## ABSTRACT

In this communication, we discuss results obtained with the numerical in-house PHP simulation code CASCO (French abbreviation for Advanced PHP simulation code). The aim is to study the impact of bubble generation (boiling) on the established oscillation of a PHP with a high turn number (10-turn water-copper tubular PHP with independently heated turns in the evaporator region). We vary the nucleation barrier value, i.e. the local superheating $\Delta T_{nucl}$ at which the nucleation occurs and the heat load $P_e$; other simulation parameters are kept constant. The PHP heat transfer performance is evaluated with the PHP thermal resistance $R_{th}$ for both horizontal and vertical favorable PHP orientation. Since the bubble generation plays a positive role for the PHP functioning, one expects that the function $R_{th}(\Delta T_{nucl})$ monotonously increases. This is true for high $\Delta T_{nucl}$. For small $\Delta T_{nucl}$, the $R_{th}(\Delta T_{nucl})$ function is decreasing, thus a minimum occurs. It appears because of a transition between the regimes of stable PHP functioning. The regimes strongly differ by the average penetration of liquid plugs into the evaporator. For horizontal PHP, these regimes correspond to the intermittent and continuous oscillations. The regimes are discussed and the regime map in the coordinates $P_e$ -$\Delta T_{nucl}$ is obtained.

The role of bubble generation (boiling) in PHP is extremely important. It has been shown [1] that, when the heat conduction via the PHP channels is accounted for and without the bubble generation, establishment of stable oscillations is impossible to achieve in general case. This occurs because the liquid plugs finally merge into a single long plug. Only small-amplitude (and thus thermally inefficient) stable oscillations might be possible under some specific conditions. The bubble generation has been first introduced into the PHP modeling by Senjaya and Inoue [2]. The state of the art bubble generation model has been developed by several other teams, see [3] for a review. The purpose of this communication is to discuss the impact of bubble generation on the stable functioning of a PHP.

The PHP simulations presented below have been performed with the in-house CASCO software (French abbreviation for Code Avancé de Simulation du Caloduc Oscillant, Advanced PHP Simulation Code) [1,3-5]. The CASCO approach is 1D. The code is stable and computationally efficient: a simulation of 300 s takes about 7 hours at a desktop PC. The PHP geometry (Fig. 1) is conventional and is the same as studied earlier [5]. The specific PHP parameters used for the simulation are listed in Table 1. The fluid is water and the tube material is copper. The long-time evolution has been analyzed: the data for 100 s$<t<$300 s (where $t$ =100 s is sufficiently large for establishment of a stable oscillation regime) were used for averaging. The thermal boundary conditions are the imposed temperature $T_c$ in condenser and the imposed heat power $P_e$ in evaporator. Each evaporator section is heated independently and homogeneously, with the power $P_e/N_{turn}$, where $N_{turn}$ is the PHP turn number. This corresponds e.g. to the heating with the electric wire wound around the tube. The PHP performance is characterized by the PHP thermal resistance $R_{th}$ $=(\langle T_e\rangle - T_c)/P_e$ , where the angle brackets mean averaging of the temperature $T_e$ of the evaporator sections of PHP channel over both time and length of these sections. $R_{th}$ is studied as a function of the most important parameter of a bubble generation model: the nucleation barrier, i.e. the local superheating $\Delta T_{nucl}$ at which nucleation occurs (see [1]).

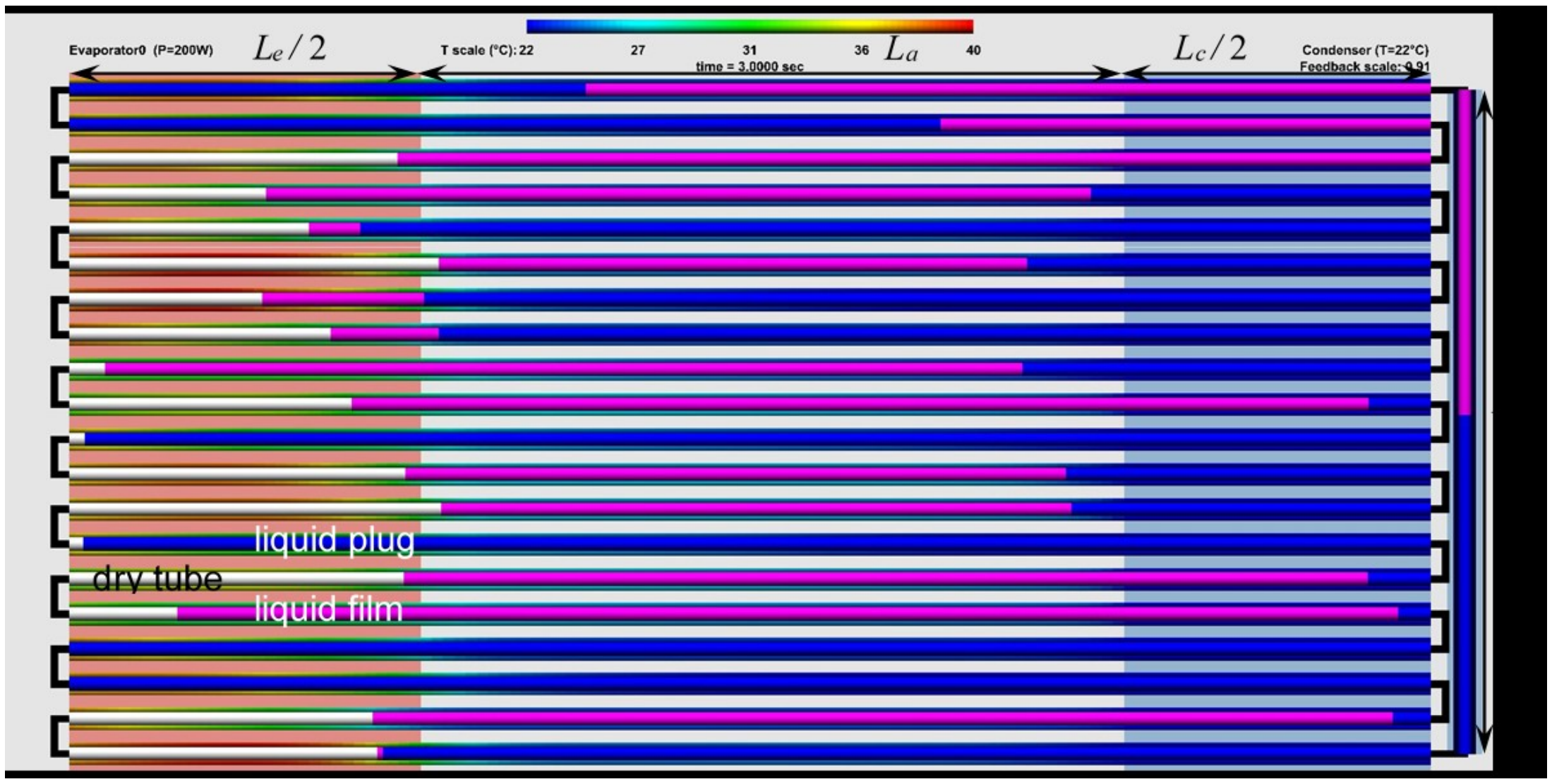

**Fig. 1** PHP geometry pictured by the CASCO data postprocessor and an example for the phase distribution inside the PHP for its horizontal orientation and large $\langle F_l \rangle$ regime. The circular elbows connecting parallel tube branches are not drawn for simplicity. The black polylines merely show the connectivity of the PHP tube; a half of elbow is included into each of the neighboring branches.

Both $P_e$ and $\Delta T_{nucl}$ are varied while other simulation parameters are kept constant. A larger $\Delta T_{nucl}$ causes a larger delay between the subsequent bubble generation events. Since the bubble generation induces the pressure imbalance inside the PHP channel, its role is beneficial for oscillations. One thus expects that the function $R_{th}(\Delta T_{nucl})$ monotonously increases. This is indeed so at large $\Delta T_{nucl}$ (Fig. 2). At small $\Delta T_{nucl}$, the function decreases so a minimum occurs. Such a minimum indicates a transition between two regimes. It is important to understand, which quantity strongly changes at this transition (in the theory of phase transitions in physics this quantity is called "order parameter"). This quantity turns out to be the fraction $\langle F_l \rangle$ [5] of evaporator covered by liquid plugs (similarly to $T_e$, averaged over time and evaporator sections). Its variation with $\Delta T_{nucl}$ is shown in Figs. 3 together with the variation of $R_{th}$. One can see a steep change of $\langle F_l \rangle$ that occurs for a value of $\Delta T_{nucl}$ slightly smaller than that of the minimum of $R_{th}(\Delta T_{nucl})$. The reason of this $\Delta T_{nucl}$ difference is a multiple steady state [6] (alternating low $\langle F_l \rangle$ and high $\langle F_l \rangle$ intervals).

**Table 1** Parameters used for the simulation.

| | |
|---|---|
| Evaporator length, $L_e$ | 126 mm |
| Length of condenser section, $L_c$ | 110 mm |
| Length of adiabatic section, $L_a$ | 126 mm |
| Number of turns, $N_{turn}$ | 10 |
| Length of the feed-back section, $L_{fb}$ | 130 mm |
| Turn curvature radius | 12.8 mm |
| Outer tube diameter | 3.2 mm |
| Inner tube diameter | 1.4 mm |
| Condenser temperature, $T_c$ | 22°C |
| Filling ratio | 0.5 |
| Nucleated bubble length | 100 μm |
| Bubble deletion threshold | 10 μm |
| Liquid film thickness | 20 μm |
| Time step | 0.1 ms |
| Tube mesh spacing | 2 mm |
| Liquid mesh spacing (approximately) | 1 mm |

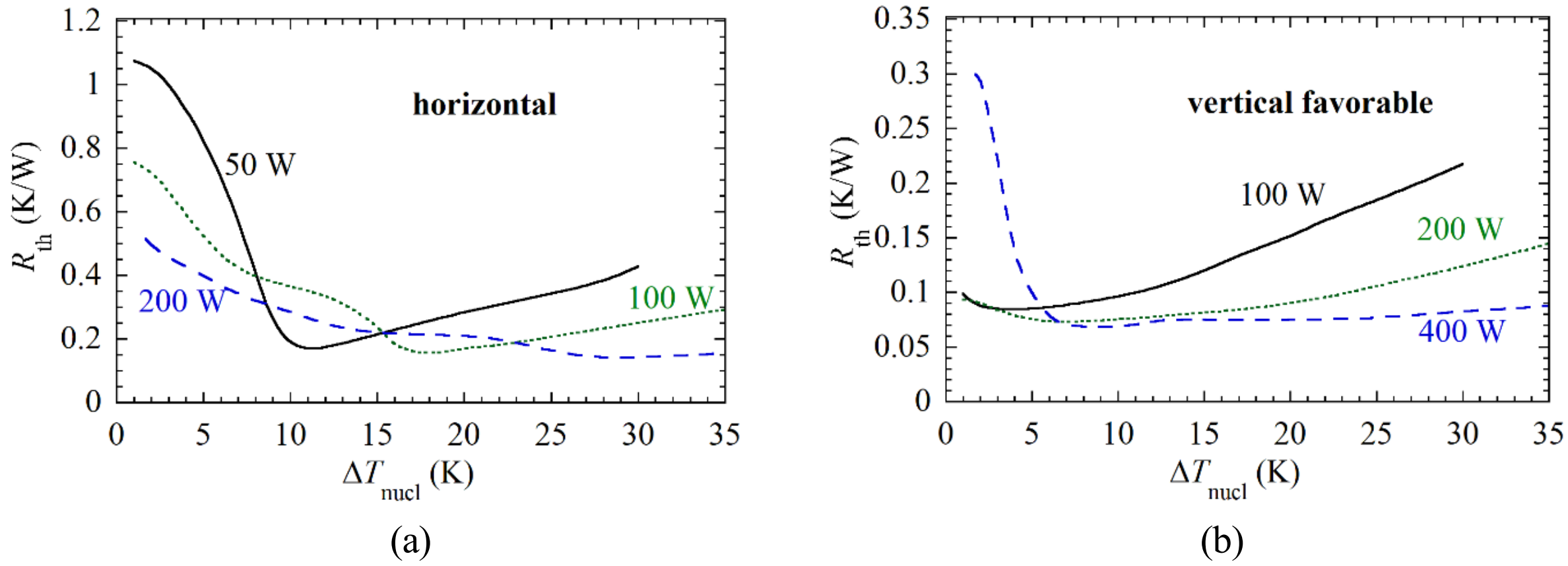

**Fig. 2** Variation $R_{th}(\Delta T_{nucl})$ at different $P_e$ for two different orientations of PHP: horizontal and vertical favorable (evaporator below condenser).

One can discuss now how the flow inside PHP changes at the transition. Consider first the case of the horizontal PHP (Figs. 2a, 3a). For small $\Delta T_{nucl}$, the flow is intermittent (or stopover) [5,6]. With the increase of $\Delta T_{nucl}$, the frequency and duration of stopovers decreases which leads to the decrease of $R_{th}$. At large $\Delta T_{nucl}$, the oscillations are continuous with no stopovers. As the efficiency of bubble generation decreases with $\Delta T_{nucl}$, $R_{th}$ increases. The jump of $\langle F_l \rangle$ is quite pronounced and the function $\langle F_l \rangle(\Delta T_{nucl})$ is close to the Heaviside function profile. The above transition can thus be interpreted as a change between intermittent and continuous oscillation regimes, see [5] for their detailed description.

The transition is less pronounced for the vertical favorable orientation of PHP (Figs. 2b, 3b). The change in $\langle F_l \rangle$ is smaller and variation of $\langle F_l \rangle$ is stronger for large $\Delta T_{nucl}$ (Fig. 3b). This occurs because the oscillations are continuous for all values of $\Delta T_{nucl}$. At small $\Delta T_{nucl}$, the oscillation mechanism is the following. Long liquid plugs oscillate mainly in the condenser and adiabatic sections. Penetration of a liquid plug into the evaporator is quite rare. It can however occur when the liquid film is dried out there (otherwise, the pressure is high because of the film evaporation and the long plug moves toward condenser rather than evaporator). Once a long liquid plug penetrates into evaporator, because of high $T_e$ and low $\Delta T_{nucl}$, a vapor bubble is generated immediately at the border between the evaporator and adiabatic sections. The bubble cuts a short liquid plug off the long plug. The bubble rapidly expands thus driving the short plug inside the dry evaporator. While moving, the trailing plug meniscus deposits the liquid film. Since the plug is short, its entire mass is spent for the film deposition before it reaches the evaporator end. The longer part of the initial liquid plug is driven back toward the condenser.

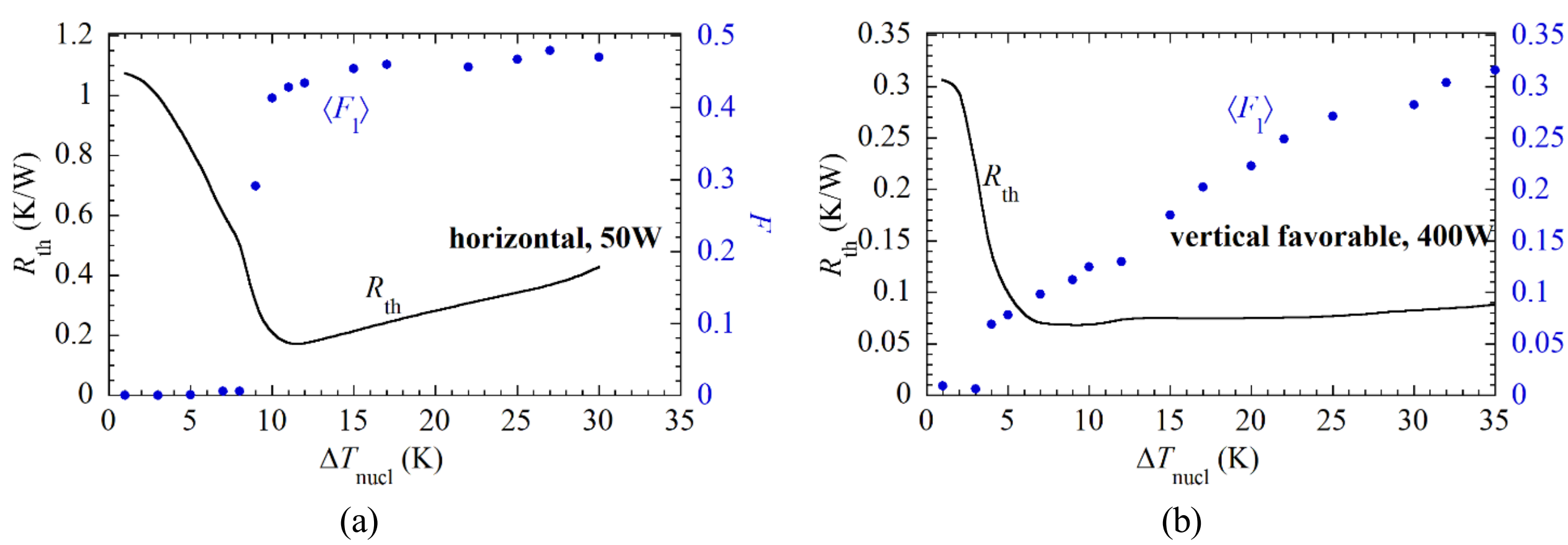

**Fig. 3** Thermal resistance $R_{th}$ (same as in Figs. 2) and average fraction $\langle F_l \rangle$ of evaporator covered by liquid plugs as functions of $\Delta T_{nucl}$ for the horizontal and vertical favorable orientations of PHP.

At large $\Delta T_{nucl}$, the oscillation mechanism is different. The bubble generation leads to formation of longer plugs that can reach adiabatic section, lead to the bubble compression and rebound back into evaporator. The plug length distribution is homogeneous (instead of being bimodal); film coverage of evaporator is larger, which causes much more efficient heat transfer.

By using the data similar to those of Figs. 2, one can obtain the regime map in the coordinates $P_e$-$\Delta T_{nucl}$ (Figs. 4). While keeping $\Delta T_{nucl}$ constant and varying $P_e$, one recovers the result obtained previously [5] for horizontal orientation: for small $P_e$ continuous oscillations occur while the intermittent regime holds for large $P_e$. Since the same kind of information is obtained experimentally for many prototypes (see e.g. [6]), one may expect that such a map applies to any PHP. Note that the average slope of the boundary is much larger for the horizontal than for vertical case.

In conclusion, the study of the impact of bubble generation on the PHP stable functioning regimes shows some important features that are expected to apply to any PHP with high turn number. There are two regimes that differ by the average fraction of evaporator coverage by liquid plugs. The regime of low coverage can be associated to the intermittent oscillation regime (if it exists). The regime of high coverage is the most advantageous for the heat transfer. It corresponds to the continuous oscillation regime. It occurs for large nucleation barriers. The obtained dependencies are important for choosing the surface properties for the PHP channels.

The authors acknowledge the financial contribution of CNES awarder through GdR MFA.

[1] I. Nekrashevych, V. S. Nikolayev, Effect of tube heat conduction on the pulsating heat pipe start-up, Appl. Therm. Eng. 117 (2017) 24 – 29.

[2] R. Senjaya and T. Inoue, Oscillating heat pipe simulation considering bubble generation. Part I: Presentation of the model and effects of a bubble generation, Int. J. Heat Mass Transf. 60 (2013) 816–824.

[3] V. Nikolayev, M. Marengo, Pulsating Heat Pipes: Basics of Functioning and Numerical Modeling, in: J. R. Thome (Ed.), Encyclopedia of Two- Phase Heat Transfer and Flow IV, vol. 1: Modeling of Two-Phase Flows and Heat Transfer, World Scientific, 2018: pp. 63 – 139.

[4] V. S. Nikolayev, A Dynamic Film Model of the Pulsating Heat Pipe, J. Heat Transfer 133 (2011) 081504.

[5] I. Nekrashevych and V. S. Nikolayev, Pulsating Heat Pipe Simulations: Impact of PHP Orientation, Microgravity Sci. Technol. (2019) doi:10.1007/s12217-019-9684-3.

[6] V. K. Karthikeyan, S. Khandekar, B. C. Pillai, and P. K. Sharma, Infrared thermography of a pulsating heat pipe: Flow regimes and multiple steady states, Appl. Therm. Eng. 62 (2014) 470–480.

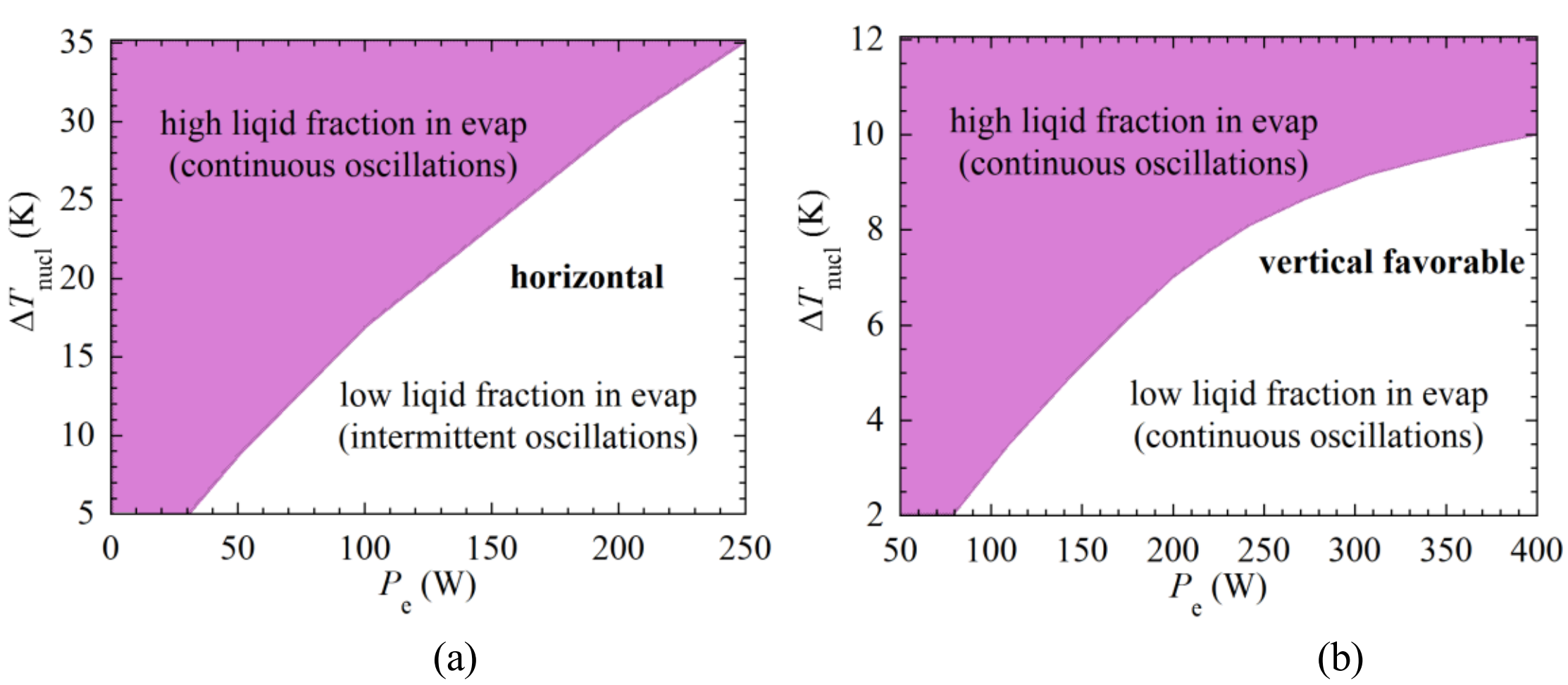

**Fig. 4** PHP regime map in the coordinates $P_e - \Delta T_{nucl}$ for the horizontal and vertical favorable orientations.